\documentclass[final,1p,times]{elsarticle} 

\usepackage{graphicx}
\usepackage{amssymb} 
\usepackage{amsmath} 
\usepackage{amsthm} 
\usepackage{lineno}
\usepackage{subfigure}

\journal{Nuclear Physics A} 

\begin{document}

\begin{frontmatter} 

\title{Constraining models of initial state with $v_2$ and $v_3$ data from LHC and RHIC}

\author[rvt]{Ekaterina Retinskaya\corref{cor1}}
\ead{ekaterina.retinskaya@cea.fr}
\author[els,els2]{Matthew Luzum}
\ead{ MWLuzum@lbl.gov }
\author[focal]{Jean-Yves Ollitrault}
\ead{jean-yves.ollitrault@cea.fr}

\cortext[cor1]{Corresponding author}

\address[rvt]{CEA, IPhT, Institut de physique th\'eorique de Saclay, F-91191
Gif-sur-Yvette, France}
\address[els]{McGill University,  3600 University Street, Montreal QC H3A 2TS, Canada}
\address[els2]{Lawrence Berkeley National Laboratory, Berkeley, CA 94720, USA}
\address[focal]{CNRS, URA2306, IPhT, Institut de physique th\'eorique de Saclay, F-91191
Gif-sur-Yvette, France}

\begin{abstract} 
We present a combined analysis of elliptic and triangular flow data from LHC and RHIC using viscous relativistic hydrodynamics. Elliptic flow $v_2$ in hydrodynamics is proportional to the participant eccentricity $\varepsilon_2$ and triangular flow is proportional to the participant triangularity $\varepsilon_3$, which means $v_n=C_n\varepsilon_n$,  where $C_n$ is the linear response coefficient in harmonic n. Experimental data for $v_2$ and $v_3$ combined with hydrodynamic calculations of $C_n$ thus provide us with the rms values of initial anisotropies $\varepsilon_2$ and $\varepsilon_3$. By varying free parameters in the hydro calculation (in particular the shear viscosity), we obtain an allowed band in the (rms $\varepsilon_2$, rms $\varepsilon_3$) plane. Comparison with Monte-Carlo models of the initial state allows us to exclude several of these models. We illustrate that the effect of changing the granularity of the initial state is similar to changing the medium properties, making these effects difficult to disentangle.

\end{abstract} 

\end{frontmatter} 


\section{Introduction}

One of the most important topics of study in heavy-ion collisions is the observation of particle momentum anisotropy in directions transverse to the beam~\cite{Voloshin:2008dg}, which can provide the evidence for the formation of some strongly interacting medium, which thermalizes and expands as a liquid, which we call the quark-gluon plasma (QGP).
The corresponding experimental observables are the flow coefficients $v_1$, $v_2$, $v_3$ etc.  In these proceedings we will concentrate on $v_2$ and $v_3$ (elliptic~\cite{Ackermann:2000tr,Aamodt:2010pa} and triangular) flow coefficients. While elliptic flow,$v_2$, is  a response of the system to an initial distribution with the form of ellipse in the transverse plane~\cite{Ollitrault:1992bk}, the triangular flow, $v_3$, is understood as the response of triangular deformation, which is caused by fluctuations of initial geometry~\cite{Alver:2010gr}.

In spite of the fact that $v_2$ and $v_3$ are the most studied harmonics of anisotropic flow,  there are still a number of open questions. Different models of initial states give different values when trying to extract  transport coefficients from data. For instance by tuning $\eta/s$ (viscosity over entropy~\cite{Kovtun:2004de}) one can match the experimental data with one model or another~\cite{Luzum:2008cw}. And it was found out that although one could fit both $v_2$ and $v_3$ data separately by tuning $\eta/s$ with hydro calculation, some of the models of initial state were unable to fit simultaneously $v_2$ and $v_3$ ~\cite{Alver:2010dn,Adare:2011tg}. This hints, that by combining $v_2$ and $v_3$ data we can constrain models of initial state even though the viscosity is unknown. 

 \section{Monte Carlo models of initial state}
 \label{s:models}
 
By initial conditions, we mean the initial energy-density profile at thermalization time $t_0$~\cite{Gelis:2013rba}. This profile is not smooth and has fluctuations from wave-functions of incoming nuclei. The magnitude of these fluctuations is still to a large extent unconstrained from data. Another open question pertaining to initial state is how elongated is the ellipse of the overlap area in non-central collisions. We will address these issues and  test different Monte Carlo models of initial state. We are testing two types of models: Glauber-type models and QCD-inspired models. The Monte Carlo Glauber model is the oldest and the most classic one~\cite{Miller:2007ri}. We use the PHOBOS Monte Carlo ~\cite{Alver:2008aq}.  In this model positions of nucleons within a nuclei are sampled through Monte Carlo. These nucleons move on straight lines and interact if their distance is less then $\sqrt{\sigma_{NN}/\pi}$. Typically one then models each nucleon as a Gaussian source, so that the final energy-density is equal to a sum of Gaussians. 

Among the QCD-inspired models we are going to test 4 of them: the oldest QCD model which we call MC-KLN~\cite{Drescher:2007ax,Albacete:2010ad}, which is using $k_T$ factorization and taking into account fluctuations of the positions of the nucleons. The second model MCrcBK~\cite{Dumitru:2012yr} is an improved MC-KLN model with additional KNO fluctuations in order to match multiplicity distribution in pp collisions. The third one is DIPSY~\cite{Flensburg:2011wx}, a QCD model which takes into account the multiple gluon cascade. And the last one is the IP Glasma model~\cite{Schenke:2012hg}, which doesn't assume $k_T$ factorization and includes non-linearities and fluctuations of color charges within a nucleon.

\section{Hydro evolution}

In spite of the fact that collaborations have published data for integrated  $v_n$\cite{Borghini:2000sa} for n=1, 2, ... 6~\cite{ATLAS:2012at}, here we only  use elliptic and triangular flow. The main reason is that these two Fourier coefficients are determined by  simple linear response to the initial state~\cite{Niemi:2012aj}. That means: $v_2\propto\varepsilon_2$~\cite{Holopainen:2010gz}, where $\varepsilon_2$ is called participant ellipticity~\cite{Alver:2006wh} and  $v_3\propto\varepsilon_3$~\cite{Petersen:2010cw}, $\varepsilon_3$ is called participant triangularity~\cite{Alver:2010gr}.  The participant eccentricity $\varepsilon_n$ for a single event is defined as~\cite{Teaney:2010vd,Bhalerao:2011yg}:
 \begin{equation}
\varepsilon_{n}=\frac{|\{r^n e^{in\phi}\}|}{\{r^n\}},
\end{equation}
where $\{...\}$ denotes an average value over the initial energy density (it can also be entropy density profile though) after recentering the coordinate system $\{re^{i\phi}\}=0$. 

Assuming linear response to the initial anisotropy, the anisotropic flow in an event is $v_n=C_n \varepsilon_n$. The response coefficient $C_n$ is the 
same for all events in a centrality bin, but $\varepsilon_n$ fluctuates, so that initial-state fluctuations result in event-by-event flow fluctuations. 
Experiments measure the rms value of $v_n$ over a centrality bin, thus the experimentally measured flow $v_n$ is proportional to the rms value of  $\varepsilon_n$. We can therefore write:
\begin{equation}
\label{rms}
\sqrt{\langle\varepsilon_{n}^2\rangle}=\frac{\sqrt{\langle(v_{n})^2\rangle}}{C_n},
\end{equation}
where $\sqrt{\langle(v_{n})^2\rangle}$  is the measured root mean square value of integrated flow and  $\langle...\rangle$ represents an average over collision events. The response coefficient $C_n$ is calculated in hydrodynamics as $C_n=(v_n/\varepsilon_n)_{hydro}$, so that we are able to extract the root mean square values of $\varepsilon_2$ and $\varepsilon_3$.
We take experimental data of integrated flow from the ALICE and PHENIX collaborations~\cite{Adare:2011tg,ALICE:2011ab}.

\section{Uncertainties in hydro response}
The standard hydro modeling~\cite{Gale:2013da} procedure consist of 3 main steps:\\
1) initial conditions\\
2) evolve these initial conditions through relativistic hydro evolution\\
3) convert the liquid into hadrons at freeze-out temperature.

Each of these steps obviously has its uncertainties~\cite{Luzum:2012wu}. The main uncertainty in the hydro evolution is the value of the shear viscosity of the strongly-interacting quark-gluon plasma. This value is not constrained well in theory and experiment\cite{Meyer:2007ic,Song:2012ua}, so we  vary $\eta/s$ as a parameter from 0 to 0.24 in steps of 0.04. The value of the shear viscosity has a remarkable influence on the values of integrated flow: the flow decreases with increasing viscosity. Another big source of uncertainty is coming from initial conditions.
Our hydro calculations are 2+1D viscous hydrodynamic, which uses as input initial condition the transverse energy density profile from an optical Glauber model. This profile is smooth and already has an ellipticity, so automatically gives us elliptic flow values. If we want to obtain the values of $\varepsilon_3$ or $v_3$, with this profile we get 0 for both of them, so in order to calculate $C_3$ we deform the third harmonic in the profile in the following way~\cite{Alver:2010dn} :

\begin{equation}
\epsilon(r, \phi) \rightarrow (r\sqrt{1+\varepsilon'_n \cos(n(\phi-\psi_n)}), \phi)
\end{equation}
where $\varepsilon_n'$ is magnitude of the deformation, and $\psi_n$ is the orientation of the deformation. 

In order to estimate the uncertainty on the value of $C_n$ from the initial profile, we use two definitions of $\varepsilon_n$: with energy density weighting and entropy density weighting.

Another free parameter is the thermalization time $t_0$\cite{Gelis:2013rba,Kolb:2003dz} which is not known. We vary it from 0.5 fm/c to 1fm/c. While we vary the thermalization time we tune the starting temperature $T_{start}$ and the freeze-out temperature $T_{fr}$ so as to match the $p_t$ spectrum \cite{Adamczyk:2013waa,Luzum:2009sb}.

In the hadronic part we also have some uncertainty, which is coming from the freezeout viscous correction, the momentum dependence of which is unknown\cite{Teaney:2003kp,Teaney:2013gca,Dusling:2009df}, so we test 2 possible ansatzs: linear $\propto p$\cite{Luzum:2010ad} and quadratic $\propto p^2$\cite{Teaney:2003kp}.
We use the same code as in Ref.[38] and we take into account resonance decays after hadronization.

\begin{figure}
\begin{minipage}{0.48\textwidth} 
\centering
\includegraphics[width=1.1\linewidth]{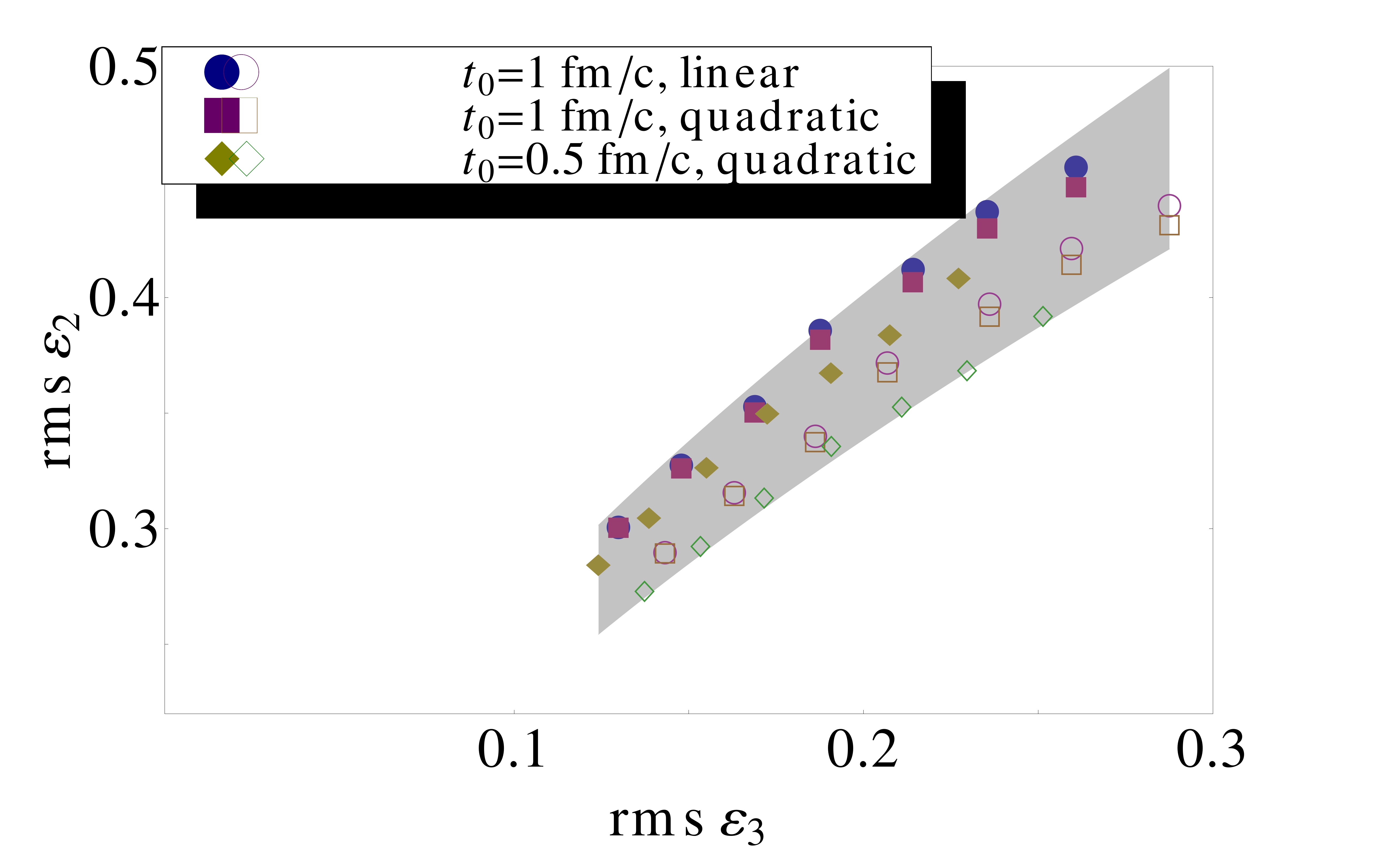}
\caption{ (Color online) R.m.s. values of $\varepsilon_2$($\varepsilon_3$)  from hydro simulations + ALICE data for  20-30\% centrality range. Purple squares correspond to $t_{init}=1fm/c$ with quadratic freezeout. Blue circles correspond to $t_{init}=1fm/c$ with linear freezeout. Yellow diamonds correspond to  $t_{init}=0.5 fm/c$ with quadratic  freezeout. Open symbols mean entropy-density profile used. The shaded band is an allowed band encompassing uncertainty in the extracted values.}
\label{fig:figure1}
\end{minipage}\hspace{0.04\textwidth}
\begin{minipage}{0.48\textwidth}
\centering
\includegraphics[width=1.1\linewidth]{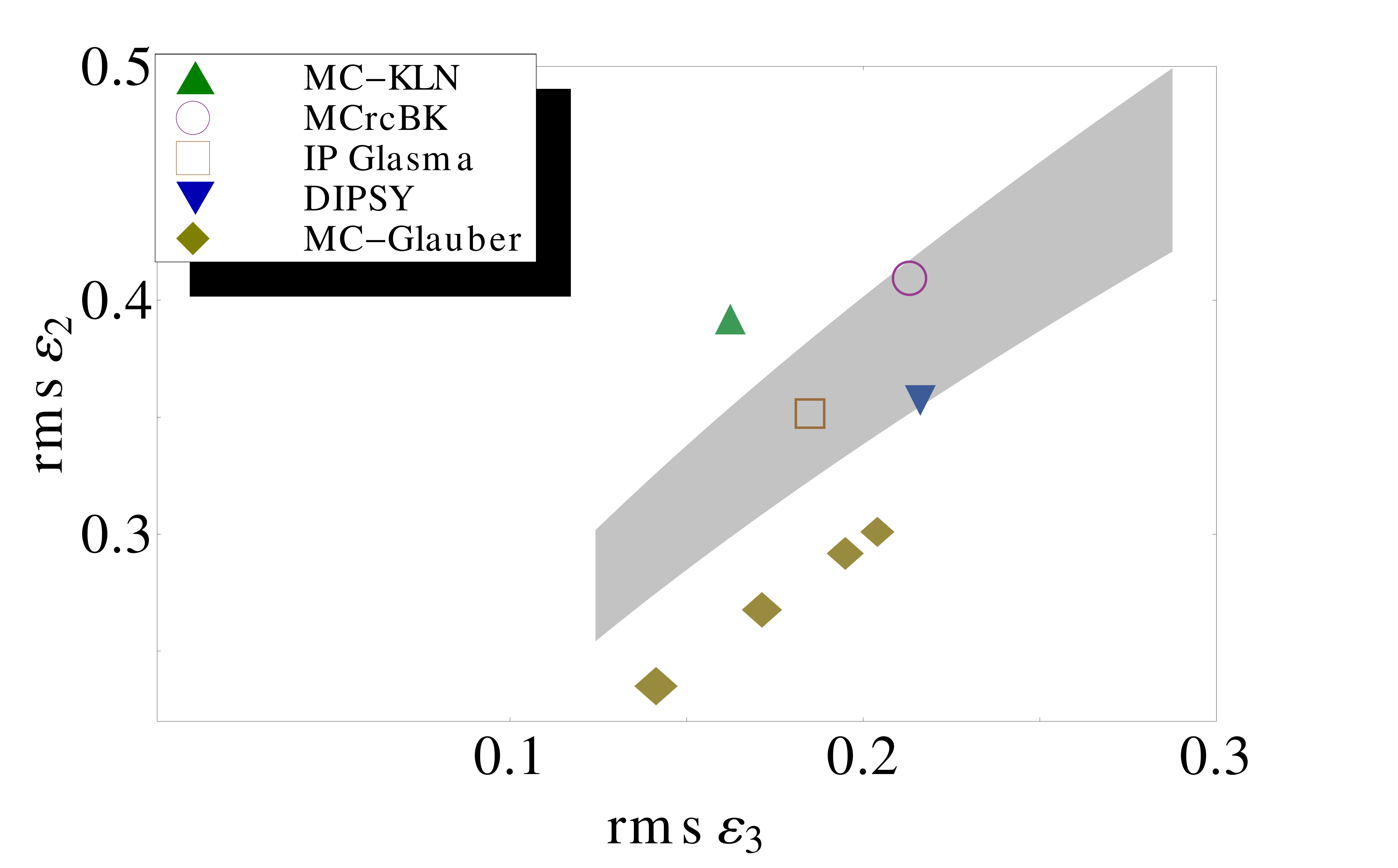}
\caption{(Color online)
The shaded band is the same as in Fig.1 and represents allowed values. Symbols are predictions from various models of initial state. The MC-Glauber model is shown for different values of the width of gaussian $\sigma$=0 fm, 0.4 fm, 0.8 fm and 1.2 fm, which are distinguished by different symbol sizes, showing that changing the smearing parameter has the same effect as changing viscosity. \protect \phantom{textextexttexttextagaintextalotoftextblablabla}}
\label{fig:figure2}
\end{minipage}
\end{figure}

After taking into account all types of uncertainties we can calculate $\varepsilon_{2}$ and $\varepsilon_{3}$ using Eq.~(\ref{rms}). The resulting values are shown in the Fig.~\ref{fig:figure1} for the most central collisions at the LHC.

Each point in this figure corresponds to one hydro calculation with one set of parameters. We have 6 types of symbols, corresponding different sets of parameters: thermalization time $t_0$, the type of ansatz and the type of $\varepsilon_n$ weighting. These symbols are composing 6 lines with viscosity $\eta/s$ changing.  Each line has 7 points, corresponding to $\eta/s$ = 0, 0.04, 0.08, 0.12, 0.16, 0.2, 0.24 (from left to right). The first line, composed of blue circles, uses as parameters the thermalization time 1fm/c and the linear ansatz.  The line with purple squares uses thermalization time 1fm/c and quadratic ansatz. We see the line almost doesn't change, except for high viscosity, which makes values of $\varepsilon_2$ slightly smaller. The line composed of yellow diamonds corresponds to thermalization time 0.5 fm/c and quadratic ansatz. We see that $\varepsilon_2$ and $\varepsilon_3$ are both decreasing. The explanation is that since we start the hydro evolution earlier, we produce more flow, and from the ratio~(\ref{rms}) obviously if we produce more flow, we will have smaller values of $\varepsilon_2$ and $\varepsilon_3$. 
Lines composed of open symbols have the same parameters, except that entropy-density weighting is used. Now we create a shaded band, such that all these points are inside this band. This band defines the allowed range. The important fact here is that even with all uncertainties taken into account, we obtain a narrow band, which eventually allows us to constrain models. 
These lines are noticed to be well fitted by the law  $\sqrt{\langle\varepsilon_2^2\rangle}/\left(\sqrt{\langle\varepsilon_3^2\rangle}\right)^k$=$\mathcal{C}$, where k=0.6 for LHC and k=0.5 for RHIC and $\mathcal{C}$ is fixed. By computing the maximum and minimum values of $\mathcal{C}$ allowed by hydrodynamics, we determine the range of allowed values for $\mathcal{C}$.


By computing the values of $\mathcal{C}$ in various Monte Carlo models, one can check if the values predicted by models are inside the allowed region, as  shown in Fig.~\ref{fig:cvalues}.  In this way the formula can be used easily by any group who has an MC model of initial states in order to see if their model is compatible with experimental data.

\section{Results}
\label{s:Results}
We calculate the allowed region in the (rms $\varepsilon_2$, rms $\varepsilon_3$) plane for different centralities for LHC and RHIC, which we represent as a shaded band. After this we test the models, introduced in Section~\ref{s:models}. 

In Fig.~\ref{fig:figure2} we display as an 
example the 20-30\% centrality range. The MC-Glauber model is shown for different values of the width of gaussian $\sigma$=0 fm, 0.4 fm, 0.8 fm and 1.2 fm, which are distinguished by different symbol sizes. By changing this parameter the result moves parallel to the band, which has the same effect as changing the 
viscosity, so that compatibility with data cannot be improved by adjusting the unknown source size.

\begin{figure}
\begin{center}
  \includegraphics[width=.495\linewidth]{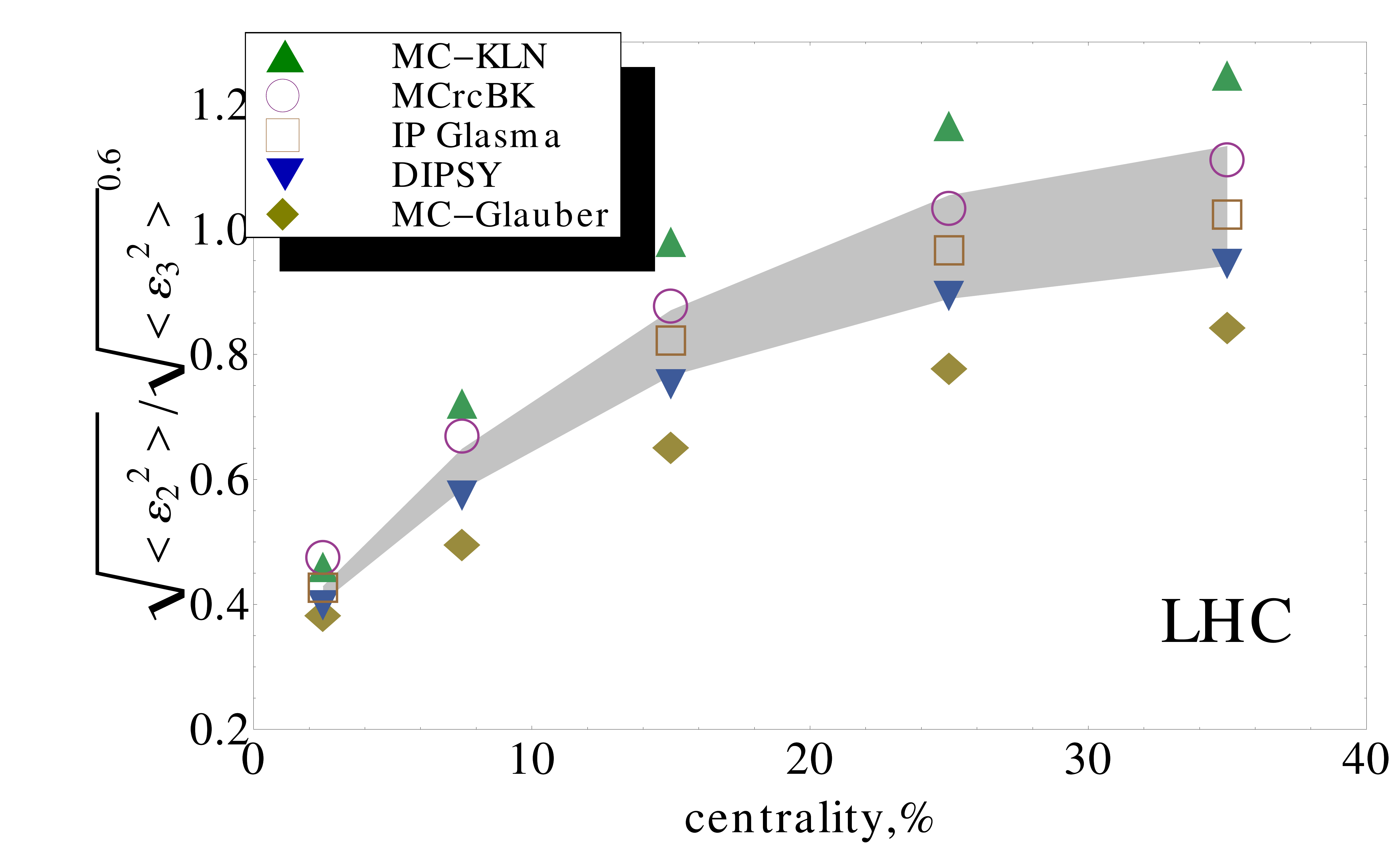}
  \includegraphics[width=.495\linewidth]{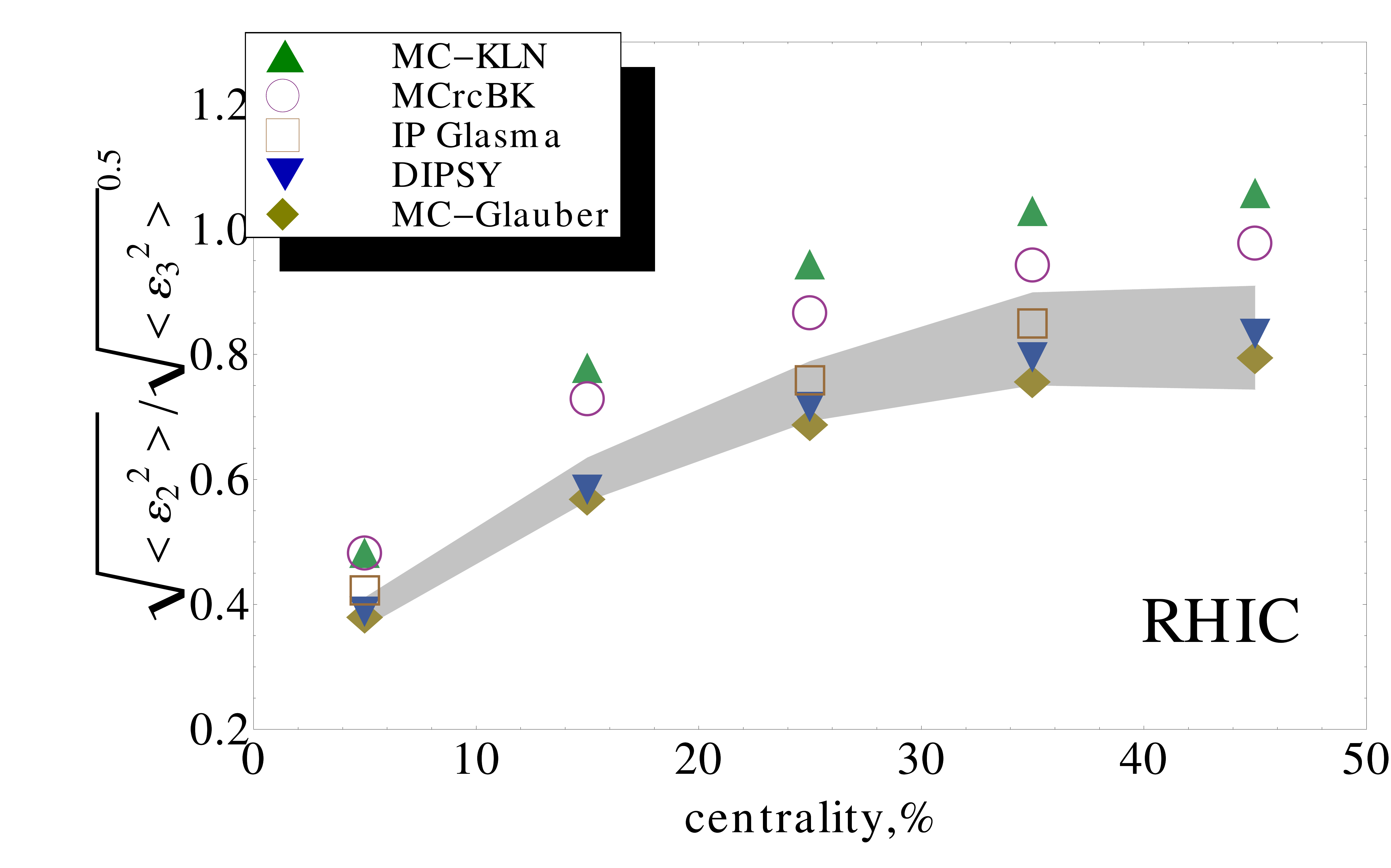}
\end{center}
\caption{(Color online) Ratio of eccentricity moments  $\sqrt{\langle\varepsilon_2^2\rangle}/\left(\sqrt{\langle\varepsilon_3^2\rangle}\right)^k$ versus centrality. Shaded bands are allowed by experiment values, combined with hydrodynamic calculations, for LHC and RHIC. Symbols are predictions from various models of initial state.}  
\label{fig:cvalues}
\end{figure}

Our main results are presented in Fig.~\ref{fig:cvalues}. It displays the rms values of $\sqrt{\langle\varepsilon_2^2\rangle}/\left(\sqrt{\langle\varepsilon_3^2\rangle}\right)^k$ versus centrality, where shaded bands are allowed values and symbols are predictions from different models. We can see that one can exclude MC-Glauber (as was also noticed for 20-30\% centrality) and MC-KLN models for LHC energies. It seems MC Glauber works better for lower energies, and MC-KLN doesn't have enough fluctuations. For RHIC energies MC-KLN can also be excluded, along with the MCrcBK model, which seems to works better at LHC energies.

From the Fig.2  in our paper~\cite{Retinskaya:2013gca}  we've seen that rms $\varepsilon_2$ and $\varepsilon_3$ values predicted by MC models both increase with centrality, but we noticed that $\varepsilon_2$ values are increasing faster then $\varepsilon_3$, which can be explained by the fact, that $v_2$ is growing faster with centrality then $v_3$: $v_2$ grows due to geometry, and $v_3$  due to the fact that fluctuations have more influence with increasing centrality, but this effect is weaker. By looking at the $\varepsilon_3$ values of MC models we can see which of the models have more fluctuations, for example by comparing MC-KLN and MCrcBK, obviously, the second has more fluctuations, as the result it has bigger value of $\varepsilon_3$.  The same about DIPSY which seems to have big value of fluctuations. 

\section*{Conclusions}
We have extracted ellipticity $\varepsilon_2$ and triangularity $\varepsilon_3$, using experimental data and hydro calculations with different sources of uncertainties and created a narrow allowed region on the (rms~$\varepsilon_3$,~rms $\varepsilon_2$) plane.
We have shown that we are able to constrain models of initial state.
It was shown that we can exclude MC-Glauber and MC-KLN models for LHC and MC-KLN and MCrcBK models for RHIC. 
We have illustrated for the MC Glauber model that changing the granularity of the initial condition model has the same effect as changing viscosity, so the effects are difficult to disentangle.

More details about this study can be found in our recently published paper~\cite{Retinskaya:2013gca}.

\bibliographystyle{h-elsevier3}
\bibliography{spires}

\begin{thebibliography}{10}

\bibitem{Voloshin:2008dg} 
  S.~A.~Voloshin, A.~M.~Poskanzer and R.~Snellings,
  arXiv:0809.2949 [nucl-ex].

\bibitem{Teaney:2010vd} 
  D.~Teaney and L.~Yan,
  Phys.\ Rev.\ C {\bf 83}, 064904 (2011)
  [arXiv:1010.1876 [nucl-th]].

\bibitem{Ackermann:2000tr} 
  K.~H.~Ackermann {\it et al.}  [STAR Collaboration],
  Phys.\ Rev.\ Lett.\  {\bf 86}, 402 (2001)
  [nucl-ex/0009011].

\bibitem{Aamodt:2010pa} 
  K. Aamodt {\it et al.}  [ALICE Collaboration],
  Phys.\ Rev.\ Lett.\  {\bf 105}, 252302 (2010)
  [arXiv:1011.3914 [nucl-ex]].

\bibitem{Ollitrault:1992bk} 
  J.~-Y.~Ollitrault,
  Phys.\ Rev.\ D {\bf 46}, 229 (1992).
  
\bibitem{Alver:2010gr} 
  B.~Alver and G.~Roland,
  Phys.\ Rev.\ C {\bf 81}, 054905 (2010)
  [Erratum-ibid.\ C {\bf 82}, 039903 (2010)]
  [arXiv:1003.0194 [nucl-th]].

\bibitem{Kovtun:2004de} 
  P.~Kovtun, D.~T.~Son and A.~O.~Starinets,
  Phys.\ Rev.\ Lett.\  {\bf 94}, 111601 (2005)
  [hep-th/0405231].

\bibitem{Luzum:2008cw} 
  M.~Luzum and P.~Romatschke,
  Phys.\ Rev.\ C {\bf 78}, 034915 (2008)
  [Erratum-ibid.\ C {\bf 79}, 039903 (2009)]
  [arXiv:0804.4015 [nucl-th]].

\bibitem{Alver:2010dn} 
  B.~H.~Alver, C.~Gombeaud, M.~Luzum and J.~-Y.~Ollitrault,
  Phys.\ Rev.\ C {\bf 82}, 034913 (2010)
  [arXiv:1007.5469 [nucl-th]].
  
\bibitem{Adare:2011tg} 
  A.~Adare {\it et al.}  [PHENIX Collaboration],
  Phys.\ Rev.\ Lett.\  {\bf 107}, 252301 (2011)
  [arXiv:1105.3928 [nucl-ex]].
  
 
\bibitem{ALICE:2011ab} 
  K.~Aamodt {\it et al.}  [ALICE Collaboration],
  Phys.\ Rev.\ Lett.\  {\bf 107}, 032301 (2011)
  [arXiv:1105.3865 [nucl-ex]].

\bibitem{Borghini:2000sa} 
  N.~Borghini, P.~M.~Dinh and J.~-Y.~Ollitrault,
  Phys.\ Rev.\ C {\bf 63}, 054906 (2001)
  [nucl-th/0007063].
  
\bibitem{Miller:2007ri} 
  M.~L.~Miller, K.~Reygers, S.~J.~Sanders and P.~Steinberg,
  Ann.\ Rev.\ Nucl.\ Part.\ Sci.\  {\bf 57}, 205 (2007)
  [nucl-ex/0701025].
 
\bibitem{Alver:2008aq} 
  B.~Alver, M.~Baker, C.~Loizides and P.~Steinberg,
  arXiv:0805.4411 [nucl-ex].

  
\bibitem{Drescher:2007ax} 
  H.~-J.~Drescher and Y.~Nara,
  Phys.\ Rev.\ C {\bf 76}, 041903 (2007)
  [arXiv:0707.0249 [nucl-th]].
  
\bibitem{Albacete:2010ad} 
  J.~L.~Albacete and A.~Dumitru,
  arXiv:1011.5161 [hep-ph].
  
\bibitem{Flensburg:2011wx} 
  C.~Flensburg,
  arXiv:1108.4862 [nucl-th].

\bibitem{Dumitru:2012yr} 
  A.~Dumitru and Y.~Nara,
  Phys.\ Rev.\ C {\bf 85}, 034907 (2012)
  [arXiv:1201.6382 [nucl-th]].
  

\bibitem{Schenke:2012hg} 
  B.~Schenke, P.~Tribedy and R.~Venugopalan,
  Phys.\ Rev.\ C {\bf 86}, 034908 (2012)
  [arXiv:1206.6805 [hep-ph]].


  

  
  

\bibitem{ATLAS:2012at} 
  G.~Aad {\it et al.}  [ATLAS Collaboration],
  Phys.\ Rev.\ C {\bf 86}, 014907 (2012)
  [arXiv:1203.3087 [hep-ex]].

\bibitem{Niemi:2012aj} 
  H.~Niemi, G.~S.~Denicol, H.~Holopainen and P.~Huovinen,
  Phys.\ Rev.\ C {\bf 87}, 054901 (2013)
  [arXiv:1212.1008 [nucl-th]].

\bibitem{Holopainen:2010gz} 
  H.~Holopainen, H.~Niemi and K.~J.~Eskola,
  Phys.\ Rev.\ C {\bf 83}, 034901 (2011)
  [arXiv:1007.0368 [hep-ph]].

\bibitem{Alver:2006wh} 
  B.~Alver {\it et al.}  [PHOBOS Collaboration],
  Phys.\ Rev.\ Lett.\  {\bf 98}, 242302 (2007)
  [nucl-ex/0610037].


\bibitem{Petersen:2010cw} 
  H.~Petersen, G.~-Y.~Qin, S.~A.~Bass and B.~Muller,
  Phys.\ Rev.\ C {\bf 82}, 041901 (2010)
  [arXiv:1008.0625 [nucl-th]].

\bibitem{Bhalerao:2011yg} 
  R.~S.~Bhalerao, M.~Luzum and J.~-Y.~Ollitrault,
  Phys.\ Rev.\ C {\bf 84}, 034910 (2011)
  [arXiv:1104.4740 [nucl-th]].

%
%
%
%

\bibitem{Gale:2013da} 
  C.~Gale, S.~Jeon and B.~Schenke,
  Int.\ J.\ Mod.\ Phys.\ A {\bf 28}, 1340011 (2013).
  

\bibitem{Luzum:2012wu} 
  M.~Luzum and J.~-Y.~Ollitrault,
  Nucl.\ Phys.\ A904-905 {\bf 2013}, 377c (2013)
  [arXiv:1210.6010 [nucl-th]].

\bibitem{Soltz:2012rk} 
  R.~A.~Soltz, I.~Garishvili, M.~Cheng, B.~Abelev, A.~Glenn, J.~Newby, L.~A.~Linden Levy and S.~Pratt,
  Phys.\ Rev.\ C {\bf 87}, 044901 (2013)
  [arXiv:1208.0897 [nucl-th]].

%


\bibitem{Adamczyk:2013waa} 
  L.~Adamczyk {\it et al.}  [STAR Collaboration],
  Phys.\  Rev.\  C 88, {\bf 014904} (2013)
  [arXiv:1301.2187 [nucl-ex]].

\bibitem{Luzum:2009sb} 
  M.~Luzum and P.~Romatschke,
  Phys.\ Rev.\ Lett.\  {\bf 103}, 262302 (2009)
  [arXiv:0901.4588 [nucl-th]].

%
  
\bibitem{Gelis:2013rba} 
  F.~Gelis and T.~Epelbaum,
  arXiv:1307.2214 [hep-ph].

\bibitem{Kolb:2003dz} 
  P.~F.~Kolb and U.~W.~Heinz,
  In *Hwa, R.C. (ed.) et al.: Quark gluon plasma* 634-714
  [nucl-th/0305084].


\bibitem{Broniowski:2008vp} 
  W.~Broniowski, M.~Chojnacki, W.~Florkowski and A.~Kisiel,
  Phys.\ Rev.\ Lett.\  {\bf 101}, 022301 (2008)
  [arXiv:0801.4361 [nucl-th]].

\bibitem{Pratt:2008qv} 
  S.~Pratt,
  Phys.\ Rev.\ Lett.\  {\bf 102}, 232301 (2009)
  [arXiv:0811.3363 [nucl-th]].

%

\bibitem{Meyer:2007ic} 
  H.~B.~Meyer,
  Phys.\ Rev.\ D {\bf 76}, 101701 (2007)
  [arXiv:0704.1801 [hep-lat]].

\bibitem{Song:2012ua} 
  H.~Song,
  Nucl.\ Phys.\ A904-905 {\bf 2013}, 114c (2013)
  [arXiv:1210.5778 [nucl-th]].

%
%
%
%
%
%
%
%
%
%
%

\bibitem{Teaney:2003kp} 
  D.~Teaney,
  Phys.\ Rev.\ C {\bf 68}, 034913 (2003)
  [nucl-th/0301099].

\bibitem{Teaney:2013gca} 
  D.~Teaney and L.~Yan,
  arXiv:1304.3753 [nucl-th].


\bibitem{Dusling:2009df} 
  K.~Dusling, G.~D.~Moore and D.~Teaney,
  Phys.\ Rev.\ C {\bf 81}, 034907 (2010)
  [arXiv:0909.0754 [nucl-th]].

\bibitem{Luzum:2010ad} 
  M.~Luzum and J.~-Y.~Ollitrault,
  Phys.\ Rev.\ C {\bf 82}, 014906 (2010)
  [arXiv:1004.2023 [nucl-th]].

%
%

   

%
%
%
%
%
%



%
%
  
\bibitem{Retinskaya:2013gca} 
  E.~Retinskaya, M.~Luzum and J.~-Y.~Ollitrault,
Phys. \ Rev.\ C {\bf89}, 014902 (2014) 
  [arXiv:1311.5339 [nucl-th]].


\end{thebibliography}

\end{document}